# Magnetic field imaging by cosmic-ray muons (Magic-μ) – First feasibility simulation for strong magnetic fields –

**Tadahiro Kin[1], Hamid Basiri[1], Eduardo Cortina Gil[2], and Andrea Giammanco[2]**

[1]Interdisciplinary Graduate School of Engineering Science, Kyushu University,
6-1 Kasuga-koen, Kasuga-shi, Fukuoka, 816-8580, Japan
[2]Centre for Cosmology, Particle Physics and Phenomenology, Université catholique de Louvain,
Chemn du Cyclotron 2, B-1348, Louvain-la-Neuve, Belgium

## Abstract

We have proposed a novel application for cosmic-ray muography, called Magic-μ, which is short for Magnetic field Imaging by Cosmic-ray Muons. The general goal of Magic-μ is to detect the presence of a magnetic field or magnetic flux density whose three-dimensional distribution is unknown. Depending on the application, Magic-μ can have three detection modes. The first is "magnetic field imaging," which is detecting the presence of a magnetic field in specific voxels within a region of space. The other two modes, transmission and deflection, aim not only to detect the presence but also to measure the flux density of the magnetic field. We have performed a feasibility study using the PHITS Monte Carlo simulation code, for strong and weak magnetic fields. In this paper, we first give an overview of the concept and basic principles of magnetic field muography. Then, the results of the feasibility study on magnetic field imaging for a strong magnetic field (more than 500 mT) are presented.

## 1. Introduction

Magnetic fields are ubiquitous, but there are very few devices or methods to measure them. For example, a Hall probe is almost the only device for measuring magnetic flux density in fixed fields. Only in cases where magnetic fields are time-dependent, can coil probing determine the number of magnetic flux variations in the loop with high temporal resolution. Measurements using these methods can be difficult for large and/or strong fields. For example, since a Hall probe can only measure a local magnetic flux density along a single axis, an enormous number of measurement points are required for a complete measurement. Therefore, actual measurements usually involve collecting data at a few measurement positions and estimating the magnetic fields at other positions by extrapolation or by making assumptions based on the symmetries of the system. This can be a source of uncertainty. Another method of representing the magnetic field is based on the distribution of iron sand around the magnet and is excellent for elementary teaching. However, images are only obtained where the iron sand is present, and this method cannot quantify the field strengths.

A two-dimensional magnetic field and its flux density were successfully determined using polarized neutrons at the Japan Proton Accelerator Research Complex (J-PARC) in Ibaraki Prefecture, Japan [1]. Using this technique, quantitative analysis of a small solenoid coil with a diameter of 5 mm and a length of about 30 mm was achieved. However, some larger magnetic fields cannot be measured because the maximum size of the target is limited by the size of the neutron beam port. In addition, special facilities are required to obtain intensely polarized neutrons. On the other hand, measurements of much stronger and spatially extended magnetic fields are desirable in cutting-edge scientific and engineering applications, such as nuclear fusion reactors and particle accelerators. Fusion reactors are a particularly important application where strong and complex magnetic fields are required to confine the plasma and sustain the fusion reactions for stable power generation. Therefore, rapid measurement of magnetic field variations is essential. Common measurement methods include a magnetic field probe array, a Rogowski coil, or a diamagnetic loop. Furthermore, many reports conclude that long-term magnetic field deviation is particularly important because the lifetime of commercial fusion reactors is primarily determined by magnetic field degradation. The cause of degradation is phase transition from superconducting material to non-superconducting material due to neutron irradiation. Currently, the degradation of the coil is monitored by supplying electric current to the superconducting coil, but the method cannot specify how and where the degradation occurs. This degradation can be revealed by the qualitative magnetic field imaging.

A plasma inspection method based on magnetic deflection of charged particles, called the heavy ion beam probe (HIBP), has been proposed and applied [2, 3] at the Texas Experimental Tokamak-Upgrade (TEXT-U), which has been equipped with a 2-MeV heavy ion accelerator for this purpose. The same concept has been successfully applied at other facilities, such as the T10 tokamak [4] and the ASDEX Upgrade tokamak . This method uses heavy-ion deflection to measure the magnetic field of the plasma. However, it cannot determine the entire magnetic field because the ions have a very weak penetrating power and it is not possible to use them inside the vessels.

To overcome this situation, we have focused on another charged particle, the muon, whose penetrating power is much greater than that of HIBP ions. Using muons, which are naturally produced by cosmic rays, is promising for probing the entire magnetic field of a large-volume, even within or behind solid materials. In practice, the "muography" images of these solid obstacles would be distorted by the magnetic field, and the pattern of distortion can in principle be used in an inversion procedure to infer the full magnetic field distribution.

Muography involves two main techniques [5]. The first, called the absorption method, was proposed by Alvarez in 1970 to search for hidden chambers in the pyramid of Khafre [6]. Recently, the absorption method has also been used to study the internal structure or density distribution of volcanoes, other pyramids, nuclear reactors, etc. Moreover, portable muography detectors have been developed for the investigation of civil infrastructures [7] or for underground exploration in confined environments [8]. Furthermore, three-dimensional imaging is also possible with this technique [9]. Since the invention of the scattering method by Borozdin et al. in 2003 [10], muography can also identify nuclides, especially heavy ones, such as uranium or plutonium. This technique is used for homeland security and nuclear waste investigations. It is based on the property of high-Z materials to scatter muons at larger angles than low-Z materials. Our proposal is a novel muography application aimed at measuring magnetic fields. We call the project Magic-μ, which is short for <u>Ma</u>gnetic field <u>I</u>maging by <u>C</u>osmic-ray <u>Mu</u>ons. This article gives a brief overview of the principles of Magic-μ and presents the results of a feasibility study for high magnetic flux densities (500 mT or more). The first simulation results for a weak magnetic field (around 200 mT) have been published by Basiri et al. [11].

## 2. Detection modes for magnetic field muography

### 2.1 Principle

For simplicity, let us assume that a single uniform magnetic field exists in a given region, as shown in Figure 1. An incident positive muon enters the magnetic field, and its trajectory is a circular motion with a Lamour radius of $r$ according to the following equation:

$$r = \frac{mv_z(0)}{B_y\sqrt{1 - \frac{v_z(0)^2}{c^2}}} \tag{1}$$

where $m, v_z(0), B_y,$ and $c$ indicate the mass of the muon, the component of the initial velocity in the z-direction, the magnetic flux density along the y-axis, and the speed of light, respectively. High-energy muons traverse the magnetic field and reach the detector position, while muons below a certain energy are reflected in the opposite direction and do not reach the detector. In this condition, the hitting position on the detector is displaced with respect to the position where it would be observed in the field-off condition. The displacement $\Delta x$ is uniquely determined by an initial muon velocity and the $BL$-value, which is defined as the magnetic flux density $\boldsymbol{B}$ multiplied by the trajectory length $L$ in the magnetic field. If we consider the same for a negative muon, we obtain the opposite displacement, $-\Delta x$. This difference is crucial for determining the direction of $\boldsymbol{B}$ because the flux at sea level is similar for positive and negative cosmic-ray muons. Therefore, we should measure the displacement along with the particle charge and velocity, event by event. Moreover, we can only roughly estimate $L$ from the incident angle $\theta$ measured by a single tracker. However, the velocity of the particles $|v|$ is not changed by magnetic fields, and the information can contribute to a more accurate estimation.

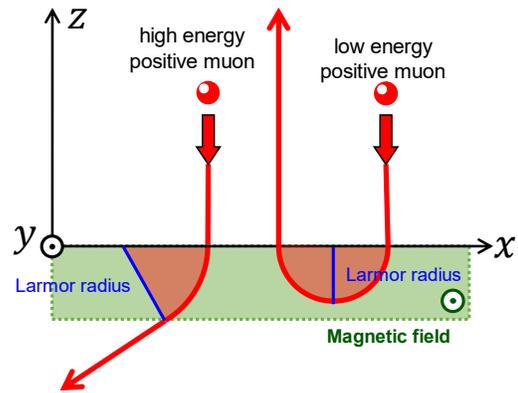

**FIGURE 1:** Low and high energy positive muons are deflected and reflected after reaction with magnetic field.

## 2.2 Magnetic field detection modes

In Magic-µ, two detection modes, qualitative and quantitative magnetic field measurements, are considered. Measurements from at least two directions (if only two, then ideally perpendicular to each other) are required for both modes to obtain 3D images. In the following, we briefly explain the characteristics of the two modes using the schematic diagram in Figure 2.

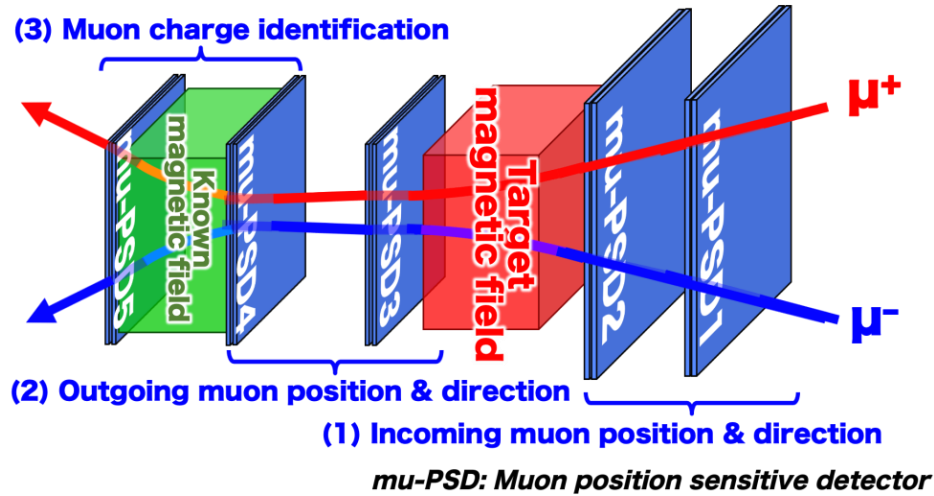

**FIGURE 2:** The two modes for qualitative and quantitative magnetic field measurements

In qualitative magnetic field measurements mode, in a similar way to absorption muography, the position and direction of the outgoing muons recorded by the detectors located downstream from the volume of interest (labelled as 2 in Fig. 2) are used to determine the presence of a magnetic field. A strong magnetic field can be considered as a reflector and return the muons back to the universe. Reflected muons never reach a muon detector located downstream of the magnetic field, similar to absorbed muons that never hit a muography detector placed downstream of the observed object. This technique can be used in many applications involving strong magnetic fields, such as the plasma control magnetic field of fusion reactors and the electric magnets of accelerators. In the next section, the simulation results of this mode for different magnetic field flux densities of ideal dipole magnetic fields are shown.

Two methods (transmission and deflection) can be used in quantitative magnetic field measurement mode. The magnetic flux density of a giant target can be derived by solving an ill-posed problem, by adding the muon charge identification part (labelled as 3 in Fig. 2) to the previous mode. This part consists of two muon position-sensitive detectors sandwiching a known magnetic field in which positive and negative muons are deflected in opposite directions. We call this method "transmission method" because it is similar to the transmission method in normal muography. The momentum of the muons is also roughly derived by combining the time-of-flight analysis and the magnitude of the deflection in the known magnetic field. This mode can be used for large-scale magnetic fields (the size of fusion reactors or accelerators).

The second method, called the "deflection method", uses the position and direction of the incoming muons (labelled as 1 in Fig. 2) to accurately determine the displacement of the muon trajectory. The detectable spatial range of the magnetic fields is smaller than in the previous modes because it is limited by the region between mu-PSDS 1,2 and 3, 4 in Fig. 2. The measurement time in this mode is similar to scattering muography (about 5 to 10 minutes) and much shorter than the other two modes. For qualitative and transmission modes, the time required for the measurement varies depending on the size and magnetic flux density of the target. For example, for targets such as fusion reactors and accelerators, our simulation results have shown that the presence of a magnetic field can be detected in a few hours.

## 3. Simulation and analysis method for first feasibility study of magnetic field imaging

### 3.1 Simulation geometry

For these simulations, we used the Monte Carlo code PHITS [12]. The cosmic-ray muons were generated with a realistic double differential flux, using the PARMA model [13, 14] in the calculation. Since Magic-µ focuses on the difference between the foreground (with magnetic field) and background (without magnetic field) muography images, we performed the simulation in two steps. First, the muography image of nine lead blocks with a height of 30 cm was obtained using absorption muography. Then, a magnetic field with 4 different magnetic field flux densities was placed 1 meter above the detector to demonstrate the effect of the magnetic field on the muography images of the lead blocks. The simulation geometry and dimensions are given in Figure 3, and we assume 152 hours of measurements for each condition.

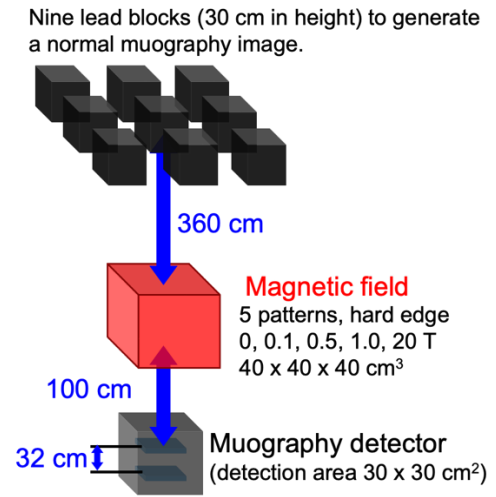

**FIGURE 3:** Simulation setup and conditions for a magnetic field imaging.

### 3.2 Analysis method

Magic-µ uses an assumption that cosmic-ray muography image is affected only by the magnetic field. In other words, the assumption states that the muography images are affected only by the absorption of objects and the deflection of muons by a magnetic field. Thus, magnetic field imaging can be found as the significance of a foreground muography image with respect to its background image. Note that unlike normal muography, the background is not an open-sky measurement but a measurement of known objects without a magnetic field. For a weak magnetic field, the effect can be observed as a distortion of the image [11]. Deconvolution analysis is required to derive the absolute magnetic flux density.

On the other hand, the scenario of a strong magnetic field is simpler, since a considerable number of muons with lower energies are deflected in a direction outside the detection range. Thus, the magnetic field acts as a momentum filter for the muons. In summary, we can straightforwardly determine the presence of a magnetic field by significance analysis of magnetic field applied image from the background image. We currently propose a simple method using threshold by a figure of merit (FOM) which is:

$$FOM = \frac{|n_{FG} - n_{BG}|}{\sigma_{FG} + \sigma_{BG}} \tag{2}$$

where $n_{FG}$, $n_{BG}$, $\sigma_{FG}$ and $\sigma_{BG}$ represent the foreground count rate, background count rate, foreground standard deviation and background standard deviation, respectively. The FOM is also expressed as FWHM ($\approx 2.35\,\sigma$). Since the FOM only indicates the degree of peak separation, we need to determine the threshold value for the magnetic field imaging and tentatively use 2 as the threshold value.

## 4. Results and discussion

The background image (muography image without a magnetic field) has clear nine lead blocks shown as "BG" in Fig.4 (a). First, we obtain a distorted image at 100 mT. Such a weak magnetic field leads to a slightly distorted image, but at stronger magnetic fields, the magnetic field acts as an energy filter for muons. Next, all muography images affected by the magnetic field were analyzed using the FOM method. When the presence of a magnetic field is detected, the pixel is colored red to create Fig.4 (b). According to the results, the regions with magnetic flux densities of 1 T or stronger magnetic fields can be detected using a threshold value of 2 for FOM. Some incorrectly detected regions were observed around the edge of the magnetic field, but this is due to the geometric conditions of the magnet. Since the magnet has hard edges, muons passing near the edges of the magnetic field escape the magnetic field, resulting in very little deflection.

Moreover, the FOM values obviously increase when the magnetic field increases. Thus, we can estimate the absolute value of the magnetic flux density (the flux direction is never determined by the magnetic field imaging mode).

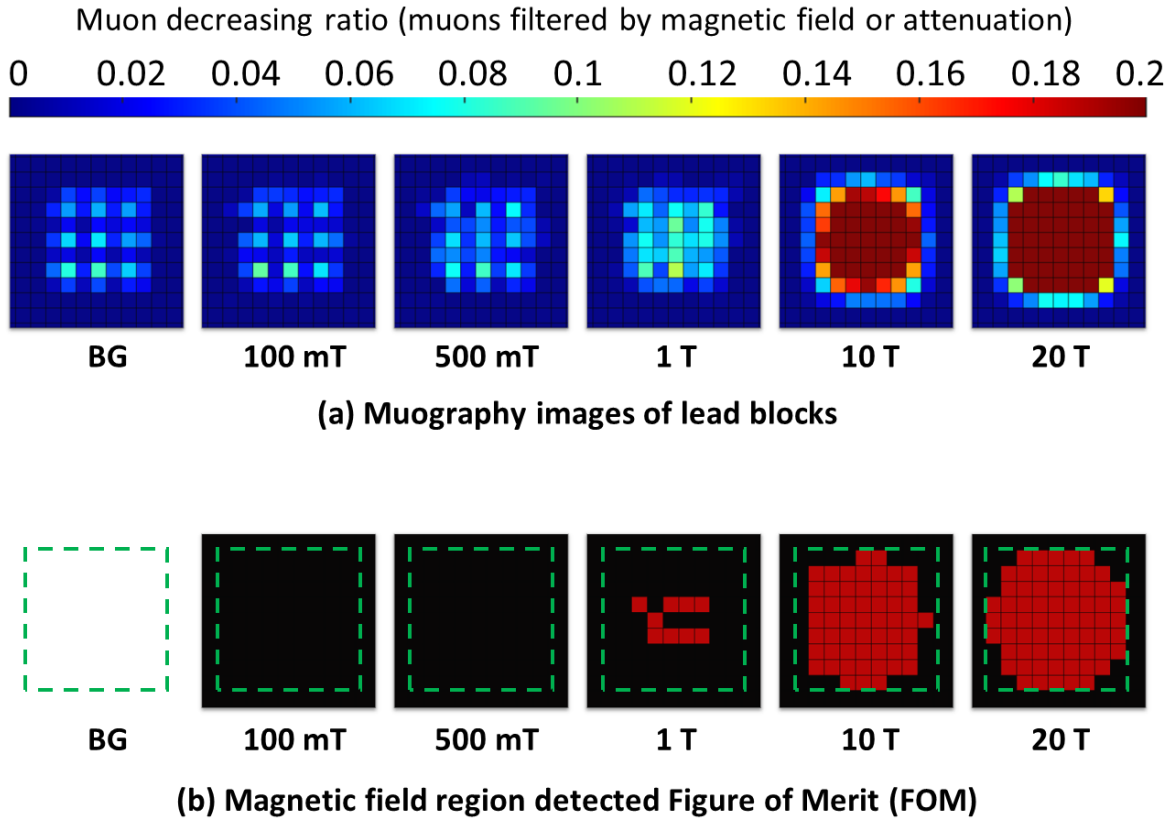

**FIGURE 4:** (a) Muography images (attenuation ratio maps) of nine blocks with magnetic field flux densities of different magnitudes. (b) Magnetic field regions determined by the FOM methods are colored red, and the actual magnetic regions is indicated by dashed lines.

To determine the measurement time, we must consider the two parameters of dimension and magnetic flux density of the target. Measurement time is defined as the time we need to find a difference between the muography image when the magnetic field is off (background image) and when the magnetic field is present. For targets similar to fusion reactors and accelerators, preliminary simulations have shown that even after a few hours of measurement for the background and foreground, the effects of the magnetic field can be seen in the muography images. In both accelerators and fusion reactors, operations are stopped for inspection at regular intervals. Therefore, it is possible to monitor the background and foreground and obtain the muography images while the magnetic field is OFF and ON.

In summary, according to the simulation results, the magnetic field imaging for nuclear fusion reactors and accelerators is promising in the range of magnetic flux density from a few to 10 T, considering their enormous size and strong magnetic field.

## 5. Conclusion

We propose a new muography application for magnetic field imaging. We call the project Magic-µ, which is short for magnetic field imaging by cosmic-cay muons. We consider three modes in the project, namely imaging, transmission, and deflection methods. This paper discusses the simulation approach of the feasibility study for one mode, namely magnetic field imaging. This objective can be achieved with any muography detector. The results demonstrated that the method is applicable for targets with enormous size and strong magnetic fields, such as fusion reactors and accelerators with magnetic flux densities of a few tesla.

## References

[1]    T. Shinohara *et al.*, Nucl. Instruments Methods Phys. Res. Sect. A Accel. Spectrometers, Detect. Assoc. Equip. **651**, 121 (2011).

[2]    T. P. Crowley, Trans. plasma Sci. **22**, 291 (1994).

[3]    D. R. Demers, P. M. Schoch, and T. P. Crowley, Phys. Plasmas **8**, 1278 (2001).

[4]    L. G. Eliseev *et al.*, Plasma Fusion Res. **13**, 10 (2018).

[5]    L. Bonechi, R. D'Alessandro, and A. Giammanco, Reviews in Physics **5**, 100038 (2020).

[6]    L. W. Alvarez *et al.*, Science (80-. ). **167**, 832 (1970).

[7]    K. Chaiwongkhot *et al.*, IEEE Trans. Nucl. Sci. **65**, 2316 (2018).

[8]    S. Wuyckens *et al.*, Philos. Trans. R. Soc. A Math. Phys. Eng. Sci. **377**, (2019).

[9]    K. Chaiwongkhot *et al.*, J. Instrum. **17**, P01009 (2022).

[10]   K. N. Borozdin *et al.*, Nature **422**, 277 (2003).

[11]   H. Basiri *et al.*, J. Adv. Instrum. Sci. **2022**, (2022).

[12]   T. Sato *et al.*, J. Nucl. Sci. Technol. **55**, 684 (2018).

[13]   T. Sato, PLoS One **10**, 1 (2015).

[14]   T. Sato, PLoS One **11**, e0160390 (2016).